\documentclass[]{agujournal_for_Arxiv}

\journalname{Water Resources Research}

\begin{document}

\title{Impact of spatially correlated pore-scale heterogeneity on drying porous media}

\authors{Oshri Borgman,\affil{1}
	Paolo Fantinel,\affil{2} Wieland L\"uhder,\affil{2}
	Lucas Goehring\affil{2,3}, and Ran Holtzman\affil{1}}

\affiliation{1}{Department of Soil and Water Sciences, The Hebrew University of Jerusalem, Rehovot 7610001, Israel}
\affiliation{2}{Max Planck Institute for Dynamics and Self-Organization (MPIDS), 37077 G\"ottingen, Germany}
\affiliation{3}{School of Science and Technology, Nottingham Trent University, Clifton Lane, Nottingham, NG11 8NS, UK}

\correspondingauthor{Ran Holtzman}{holtzman.ran@mail.huji.ac.il}




%
%
%
%
%

\paragraph*{}

\begin{abstract}
We study the effect of spatially-correlated heterogeneity on isothermal drying of porous media. We combine a minimal pore-scale model with microfluidic experiments with the same pore geometry. Our simulated drying behavior compare favorably with experiments, considering the large sensitivity of the emergent behavior to the uncertainty associated with even small manufacturing errors. We show that increasing the correlation length in particle sizes promotes preferential drying of clusters of large pores, prolonging liquid connectivity and surface wetness and thus higher drying rates for longer periods. Our findings improve our quantitative understanding of how pore-scale heterogeneity impacts drying, which plays a role in a wide range of processes ranging from fuel cells to curing of paints and cements to global budgets of energy, water and solutes in soils.
\end{abstract}

\section{Introduction} \label{sec:Intro}

Drying of porous media plays a crucial role in many natural and industrial systems, from soils, to curing of cement, paints and food~\citep{Prat2011,Goehring2015}. Drying in soils is of particular environmental importance, as it controls the transfer of water and energy between the subsurface and the atmosphere and affects solute distribution in the root zone~\citep{Or2013a}. 
The rate and extent of fluid transport in general, and drying in particular, intimately depend on the heterogenous distribution of grain and pore sizes and their connectivity, among other factors~\citep{Bultreys2016, Holtzman2016}. 
Pore-size heterogeneity, either random (disordered) or spatially-correlated in the form of patches or layers of finer or coarser particles, is an inherent property of natural porous media such as soils or sediments~\citep{Knackstedt2001}, resulting from deposition and diagenetic processes.
Even in engineered systems such as micromodels, heterogeneity is inevitable due to manufacturing errors.




In this paper, we investigate how spatial correlation in particle sizes affects isothermal drying of porous media, where evaporation is driven by vapor concentration differences between the medium and the outside atmosphere.  
Evaporation reduces the liquid pressure, allowing air to invade into the pores once the capillary entry thresholds are exceeded, forming an interface separating liquid- and air-filled pores~\citep{Lehmann2008, Shokri2010a}. 
The drying process is typically divided into two main stages. 
During stage 1 (also called the ``constant rate period'', CRP), evaporation occurs mostly from wet patches at the medium's surface, while liquid is supplied by continuous pathways from the medium's interior. Stage 1 is characterized by a drying rate which remains fairly constant despite the decline in surface wetness; this is attributed to vapor transport in an air boundary layer that develops outside the medium~\citep{Lehmann2008, Shokri2010a, Shahraeeni2012}. 
Stage 2 (or the ``falling-rate period'') is marked by a disruption of the liquid pathways to the surface, forcing the surface pores to dry out and the evaporation to occur further away from the surface. Consequently, the rate, which becomes limited by vapor diffusion within the porous medium, drops continuously~\citep{Lehmann2008,Shokri2010a,Goehring2015}.
Liquid connectivity to the surface can be further enhanced by liquid films, in systems such as throat networks with channels of noncircular cross section or with rough walls~\citep{Laurindo1998}, and granular media~\citep{Yiotis2012}. 



Previous studies have shown that when a sharp contrast in pore sizes exists (in the form of a coarser and a finer region), the coarser region dries up completely before finer pores starts to dry. This has been demonstrated experimentally for a variety of media of different length scales, including sands~\citep{Lehmann2009, Nachshon2011a,Assouline2014}, micromodels~\citep{Pillai2009} and colloidal drops~\citep{Xu2008}. 
Theoretically it has been shown that increasing the width of the pore-size distribution prolongs stage 1 by maintaining liquid connectivity for longer periods~\citep{Metzger2005,Lehmann2008}. We note that this observation is true only when gravitational or viscous forces are sufficiently large, e.g. in a sufficiently deep sample.
Nonetheless, a systematic study of how the correlation length of grain or particle sizes (defined as the characteristic length scale over which particles of similar sizes are expected) affects drying is lacking.

For the more general problem of immiscible fluid-fluid displacement, it has been shown that increasing correlation length decreases the residual saturation of the wetting phase at breakthrough~\citep{Ioannidis1993,Knackstedt2001}, leads to a more gradually-varying capillary pressure-saturation relation (retention)~\citep{Rajaram1997, Mani1999}, and improves connectivity and hence relative permeability of both phases~\citep{Mani1999}. 
Changes in fluid retention were also observed upon varying the correlation length of particle wettability in a bead pack~\citep{Murison2014}.

%
%





Here, we present a systematic investigation of the impact of spatial correlations in particle size on the drying rate and patterns in porous media. We use pore network modeling complemented with microfluidic experiments to obtain a rigorous quantitative analysis and improve our understanding of the underlying mechanisms. 
We show that increasing the correlation length promotes preferential invasion, hence preserving liquid connectivity and surface wetness, delaying the transition between the drying stages. 

\section{Methods} \label{sec:Methods}

One of the main challenges in studying fluid displacement processes such as drying is their large sensitivity to pore-scale details~\citep{Bultreys2016}. This sensitivity typically requires multiple realizations for each set of conditions in order to obtain a statistically-representative description~\citep{Mani1999}.
Here, we formulate a minimal model which describes the essential pore-scale physics of this process---including evaporation from interfaces, vapor diffusion, and capillary invasion---via a set of simple rules for interactions between pores. We use this model in computer simulations to generate a sufficiently-large data set to overcome this sensitivity. We further validate our model using microfluidic devices with state of the art manufacturing and measurement resolution.

As we seek fundamental understanding of the underlying mechanisms of drying in heterogeneous porous media, rather than an accurate quantitative description of specific materials, we choose a simplified analog of a disordered porous medium as our model system. 
In particular, we use an array of cylindrical solid pillars (our particles) placed on a regular square lattice, where heterogeneity is achieved by varying the pillar radii (Fig.~\ref{fig:ModelSchem}). 
We note that both our experimental and numerical methodologies allow use of other designs, such as a triangular lattice or a random close packing. 
%
%
We consider here a horizontal sample, to avoid gravitational effects. 
We also do not model liquid films, since for our pore geometry--unlike systems such as network of channels or granular media--such films are not expected to be well-connected (further discussion and supporting evidence are provided in Section~\ref{sec:Discussion_films}). 
%
The sample is open to the atmosphere at one of its faces, from which vapor diffuses outside and air invades into the medium (Fig.~\ref{fig:ModelSchem}a).
%
%
Our model is described below, together with a brief description of our microfluidic experiments. 
Further details of our experiments and a discussion of the ability of such a minimal model to capture the experimental behavior appear in a companion paper,~\citet{Fantinel2016}.
%

\subsection{Pore Network Model} \label{sec:MethodsPNM}

\begin{figure}
	\centering
	\includegraphics[width=.95\columnwidth]{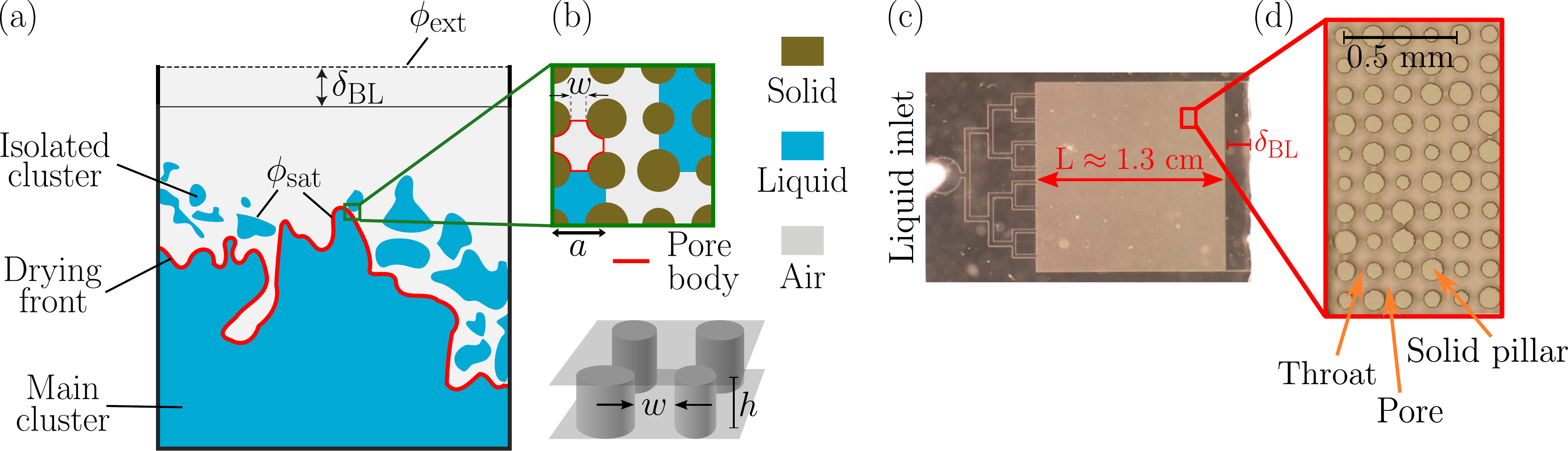}
	\caption{(a) Schematics of our model of isothermal drying of a horizontal porous sample open to the atmosphere at one of its faces. Low vapor concentration ($\phi$~$=$~$\phi_\mathrm{ext}$) at the edge of a diffusive boundary layer (of thickness $\delta_\mathrm{BL}$) that develops outside the medium drives evaporation and vapor diffusion away from air-liquid interfaces (where $\phi$~$=$~$\phi_\mathrm{sat}$). 
	Evaporation reduces liquid pressure, causing air to invade into liquid-filled pores. Some of the liquid disconnects from the bulk (i.e. the main cluster) and becomes isolated.
	The drying front, namely the leading part of the air-liquid interface (excluding isolated clusters), is marked in red.  
	(b) We model evaporation, vapor diffusion and capillary invasion via a pore network model. Pores are defined as the space (delimited by red line) between solid pillars (height $h$ and variable radius $R$) placed on a square lattice (spacing $a$), and are interconnected by throats (aperture $w$).
	(c) Drying experiments are performed in a horizontal microfluidic cell, initially filled with a volatile liquid via inlet channels. A boundary layer forms within an open region (pillar-free) of thickness $\delta_\mathrm{BL}$. (d) Close-up image showing the solid cylindrical pillars (top view).
		%
	%
	%
}
	\label{fig:ModelSchem}
\end{figure}

We develop a pore-network model of a drying porous medium, discretizing the pore space into individual pores (the space between four neighboring pillars) connected by throats (the constrictions between two adjacent pillars; Fig.~\ref{fig:ModelSchem}b); for a review of pore network models of drying see e.g.~\citet{Prat2002,Prat2011}. 
%
We extend our pore-network to include vapor diffusion in the air boundary layer above the medium's open surface, by discretizing this region into interconnected ``cells'' as in~\citet{Laurindo1998}. This layer represents the atmospheric demand of vapor, which sets the potential rate of evaporation from the porous media. We capture the two-dimensional (2-D) distribution of vapor concentration which develops in the boundary layer as the surface dries~\citep{Shahraeeni2012} (for further details see Supporting Information).
%

The evaporation rate is computed from the rate of vapor diffusion away from air-liquid interfaces. This assumption allows us to treat the interface as a source term for vapor, and compute vapor concentrations in the entire domain---air-filled pores as well as the boundary layer---by resolving the mass balance of vapor.
%
%
A further simplification is provided by the much longer timescale of vapor diffusion, relative to the timescales of interface advancement and filling of a newly invaded pore with vapor. Based on the separation of these timescales, we can represent the dynamics of the air-liquid interface as a sequence of steady-state configurations, excluding the transient evolution of vapor concentration following interface advancement from our model. With this, vapor concentrations are obtained by enforcing the continuity equation in each pore (or boundary layer cell) $i$,
\begin{equation} \label{eq:PoreMassBal}
\sum_{j} J_{ij} A_{ij}=0
\end{equation}
where the summation is done over all neighboring pores (or cells) $j$. Here 
\begin{equation} \label{eq:FickPNM}
	J_{ij} =  \left. -\rho_v^\mathrm{sat} D {\nabla \phi}\right|_{ij}
\end{equation}
is the vapor mass flux between two adjacent pores $i$ and $j$, driven by the local gradient of relative vapor concentration, $\left.{\nabla \phi}\right|_{ij} = (\phi_j-\phi_i)/l_{ij}$, where $\phi$=$\rho_v / \rho_v^\mathrm{sat}$ is the local vapor density $\rho_v$ normalized by the saturated vapor density $\rho_v^\mathrm{sat}$, and $D$ is the binary diffusion coefficient of vapor in air. Throat $ij$, connecting pores $i$ and $j$, has an effective cross-sectional area of $A_{ij} = \alpha w_{ij} h$, where $w_{ij} = a-R_{I}-R_{J}$ ($R_{I}$ and $R_{J}$ are the radii of pillar of that throat, i.e. for boundary layer cells $w_{ij}=a$) is its aperture and $h$ is the pillar height (the out-of-plane sample thickness). 
The coefficient $\alpha$ accounts for the variable pore width (varying between $w$ at the throat and $a$ at the pore's center); a value of $ \alpha=1.6$ was determined for our pore geometry by finite-element simulations at the sub-pore scale. In the boundary layer, the effective aperture corresponds to the full width of the cell, i.e. $\alpha=1$. 
The distance $l_{ij}$ is taken to be the lattice spacing $a$ if both pores are air-filled, and in boundary layer cells. 
For an air-filled pore $i$ along the air-liquid interface, we set $\phi_j=\phi_\mathrm{sat}=1$ as a boundary condition for the diffusion problem, where $J_{ij}$ represents the local evaporation rate from interface $ij$, with $l_{ij}=a/2$. Other boundary conditions are fixed concentration, $\phi = \phi_\mathrm{ext} = 0$, at the external edge of the diffusive boundary layer, and no-flux ($J=0$) conditions at all cell faces not open to the atmosphere. This provides a set of coupled linear equations in terms of $\phi$ in each pore and boundary layer cell. 
%

%

Invasion of air into liquid-filled pores depends on the local capillary pressure, or equivalently, the meniscus curvature (where the two are related via the Young-Laplace law). 
We relate the change in curvature to changes in liquid volume by approximating each throat as a cylindrical capillary tube with with an effective radius of $r_{ij}^* = \left( 1/h +  1/w_{ij} \right) ^{-1}$, with a spherical meniscus of curvature $C$, such that invasion would occur once the critical curvature for throat ${ij}$, $C_{ij}^* = 2/ r_{ij}^*$, has been exceeded, $C \geq C_{ij}^*$.
%
This approximation of $C_{ij}^* $ is $\sim$$5\%$ higher than that calculated from the Mayer-Stowe and Princen method~\citep[Eq. 55]{Lago2001}, a minor difference that should not appreciably affect our results. 
Consequently, the curvature of a meniscus $C$ is linked to the liquid volume evaporated from it, $\Delta V_{ij}$, by
\begin{equation} \label{eq:SpherCapVol}
\Delta V_{ij} = \frac{\pi (2/C)^3}{3}\left(1-\sqrt{1-\left({C}/{C_{ij}^*}\right)^2}\right)^2 \left(2+\sqrt{1-\left( {C}/{C_{ij}^*}\right)^2}\right).
\end{equation}
where $\Delta V_{ij}$ is the total volume decrement relative to a flat meniscus (where $C=0$), which is equal to the sum of the incremental changes in the volume of meniscus $ij$ prior to the current time. The incremental volume change for a time step $\Delta t$ is $\Delta t J_{ij} A_{ij}/ \rho_l$, where $\rho_l$ is the liquid density. For further details of the derivation of Eq.~(\ref{eq:SpherCapVol}) see the Supporting Information.

To resolve the curvatures of all menisci from the evaporated volume, we use the following closure relations: (1) the total liquid volume decrement from any cluster equals the sum of volumes decreased from all menisci $ij$ in that cluster, $\Delta V_{tot}=\sum_{ij} \Delta V_{ij}$; and (2) this deficit is divided between the cluster's menisci such that their curvature $C$ remains uniform (for all throats). The latter is justified by the much faster pressure diffusion in liquid than of vapor diffusion in air, allowing us to consider instantaneous liquid pressure equilibration. 
%
%
Once a pore is invaded, the liquid volume associated with it is redistributed to other interfacial pores in that cluster, decreasing the cluster's curvature $C$ according to Eq.~(\ref{eq:SpherCapVol}). Since $C$ is uniform, every meniscus can receive a different volume (according to the corresponding throat radius, $r_{ij}^*$). 
We restrict the volume of liquid that can be redistributed by enforcing $C \geq 0$; we consider a pore completely dry and advance the interface only once all of its liquid is either redistributed, or, if limited by $C \geq 0$, evaporated. 

Simulations begin with a liquid-saturated sample. Our computational algorithm is as follows: (i) The evaporation rates from the air-liquid interfaces, for a given interface configuration (invasion pattern), are computed from Eq.~(\ref{eq:PoreMassBal}); (ii) The time-step until the next invasion event is calculated. This is the time required to reach the minimal evaporated volume ($\Delta V_{tot}$) corresponding to a meniscus curvature sufficient to invade a throat, determined via Eq.~(\ref{eq:SpherCapVol}); 
(iii) The interface configuration is updated once the invaded pore empties completely (instantaneously, unless redistribution is restricted). The process is then repeated by returning to step (i), until we reach breakthrough or a desired saturation. 

\subsection{Generating Correlated Geometries} \label{sec:MethodsGeom}

%
%
%
%

For both experiments and simulations we generate samples in which the solid pillar sizes are locally correlated, that is where small pillars are more likely to be found next to other small pillars, and vice versa. The samples were generated according to the following protocol: we construct a random Gaussian surface with a prescribed spatial correlation length $\zeta$ (measured in units of the lattice spacing $a$). We then sampled this surface on the grid of pillars, and used these values to determine the pillar sizes.

In particular, the random rough surface $H(x,y)$ was generated by noting that its Fourier transform should be a Gaussian distribution of intensities, centered around zero, with random phases. This Gaussian distribution was prepared by summing a thousand sine waves, whose amplitude, phase, and orientation were selected from a random uniform distribution, and whose wave numbers were drawn from a normal distribution. The width of this distribution, in Fourier space, is inversely proportional to the correlation length $\zeta$ of the surface. For a review of methods to generate rough surfaces see~\citet{Persson2005}. We then transform this from a normal to a uniform distribution by mapping the cumulative distribution function of the random rough surface
onto the range $H(x,y) \in [-\lambda,+\lambda]$
where $\lambda$ is a measure of the heterogeneity in pillar sizes. Finally, the radius $R_I$ of each pillar is chosen as $\overline{R}(1 + H_I)$, where $H_I = H(x_I,y_I)$ is the height of the surface at the center of that pillar (coordinates $x_I,y_I$). Overbar denotes an arithmetic average throughout the text.

The following parameter values were used in both simulations and experiments: sample size of $100\times100$ pillars, with $\overline{R} = 50$ $\mu$m, $a = 130$ $\mu$m, $\overline{w} = 30$ $\mu$m, and $h = 40$ $\mu$m, providing a mean porosity of $0.53$. For the completely uncorrelated samples ($\zeta=0$), slightly different values of $a = 125$ $\mu$m and $\overline{w} = 25$ $\mu$m were used.
%
%
We generated two sets of data for this paper: (i) a large set for statistical analysis (simulations only); and (ii) a smaller set for comparison between simulations and experiments.  
Statistics including ensemble averages and deviations were obtained from a set of 40 realizations (namely samples with a different random seeds) for each $\zeta$, with $\zeta$ values of 0, 1, 1.5, 2, 2.5, 3, 4, 6, 10, and 15 (a total of 400 simulations), $\lambda=0.1$, and $\delta_\mathrm{BL}$~=~2~mm. 
Comparison of simulations and experiments was performed in samples with identical pore geometry, using 8 correlated samples ($\zeta$ values of 1, 4, 10, and 15, normally-distributed sizes with $\lambda$ values of 0.1 and 0.2) and 8 uncorrelated samples ($\zeta$=0, and $\lambda$ values of 0.03, 0.05, 0.1 and 0.2). 
%
%
%
Due to the formation of sample-spanning patches for $\zeta=15$, potentially introducing sample-scale effects, we do not include these simulations in the quantitative analysis along with other $\zeta$ values. For completeness, comparison between experimental and simulated patterns for all $\zeta$ values, including $\zeta=15$, is provided in the Supporting Information.


\subsection{Microfluidic Experiments} \label{sec:MethodsExp}

Micromodels made of an array of cylindrical pillars in between two planar plates are manufactured using standard microfluidic (aka ``lab-on-a-chip'') techniques including soft lithography. For further details see \citet{Fantinel2016}.
Briefly, a silicon wafer is spin-coated with a negative photoresist (SU8 3025, MicroChem Corp.), which is then exposed to UV light through a mask to produce the desired design (pillar positions and sizes). After rinsing, the remaining SU8 structure is used as a primer for a secondary mold of polydimethilsyloxane (PDMS). The PDMS is then cured and is used as a mold for the final sample, made of Norland Optical Adhesive 81 (NOA, Sigma-Aldrich). The NOA sample is then cured and exposed to white light to stabilize its optical properties. 
Our manufacturing procedure resolution is $\sim$2 $\mu$m, with an estimated uncertainty in pillar sizes of $\sim$1.6 $\mu$m ($\sim$3.2\% of design). The potential impact of such uncertainty is discussed in Section~\ref{sec:ResultsExpSimCompar}.

A boundary layer is included in the experimental design by leaving the region adjacent to the open side empty of solid pillars (Fig.~\ref{fig:ModelSchem}c).
%
%
This boundary layer is likely enhanced by the presence of a stagnant layer of air immediately outside of the open edge of our cell.
Thus, to compare rates we use in the simulations an effective value for $\delta_\mathrm{BL}$, computed by matching the initial experimental rates, as in~\citet{Vorhauer2015}.
A sensitivity analysis showing how uncertainty in parameters such as $\delta_\mathrm{BL}$, $D$, $\rho_v^\mathrm{sat}$ and $\overline{w} / \overline{R}$ impacts drying appears in~\citet{Fantinel2016}. 
%

The sample is initially filled with a fluorinated oil Novec 7500~\citep{3M} through inlet channels (Fig.~\ref{fig:ModelSchem}c).
At the experimental temperature (25$\pm1^{\circ}$ C), the fluid properties are: vapor pressure of 2.1$\cdot$10$^3$ Pa, $\rho_v^\mathrm{sat}$ = 0.35 kg/m$^3$, interfacial tension of $\gamma$ =
0.0162 N/m~\citep{3M}, and $D=5\cdot10^{-6}$ m$^2$/s~\citep{USEPA}. 
The cell, placed horizontally to avoid gravitational effects, is left to dry under a digital SLR camera (Nikon D5100) with a macro lens. Illumination is provided by a ring of LEDs surrounding the cell. Time-lapse images are taken every minute, with a spatial pixel resolution of 5 $\mu$m, equivalent to a liquid mass on the order of 1 nanogram (providing here a more accurate method for monitoring drying progression over weighting). Experiments are stopped at breakthrough, since past breakthrough air can invade the inlet channels.
%
%

Our image analysis procedure is briefly described below (see \citet{Fantinel2016} for further details).
First, the red color channel of the image, which contains the best contrast, is extracted. Then, a bandpass filter is applied to remove both the high-frequency noise and any low frequency variations in intensity. Subtraction of the first image from each image sequence removes constant sources of background noise. Thresholding then provides a binary image with wet (black) and dry (white) areas. Finally, we remove the solid pillars to obtain a continuous distribution of liquid and air. 
To compare with simulations we also generate a discrete data set of the invasion time for each pore by identifying the locations of the pores and their occupancy. 
%
%
%

\section{Results} \label{sec:Results}

\subsection{Comparing Model with Experiments} \label{sec:ResultsExpSimCompar}

\subsubsection{Drying Patterns} 

\begin{figure}[]
	\centering
	\includegraphics[width=.95\columnwidth]{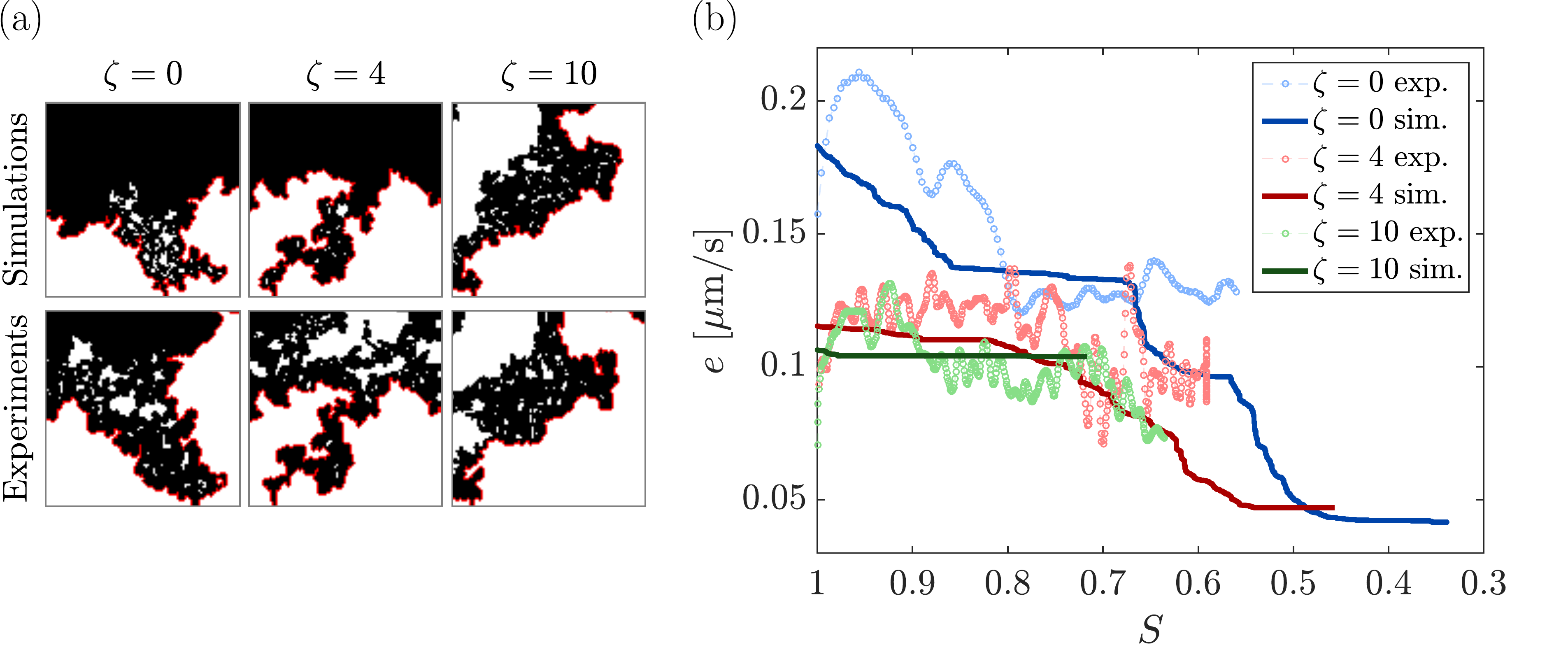}
	\caption{(a) Simulated and experimental patterns for samples with different correlation length, $\zeta$, are in good agreement. The main difference is the persistence of near-surface isolated clusters for longer periods experimentally. Black and white denote pores filled with air and liquid, respectively (solid not shown); the red line indicates the drying front. (b) Simulated drying rates $e$ initially follow the experimental ones, however they show a faster rate drop at lower liquid saturation, $S$. This drop is associated with reduced persistence of isolated clusters in simulations, pushing the drying front deeper, and thus lowering the drying rate.}
	\label{fig:ExpSimComp}
\end{figure}

Our simulated drying patterns agree well with the corresponding patterns in microfluidic experiments using identical pore geometries (Fig.~\ref{fig:ExpSimComp}a; additional patterns are provided as Supporting Information). The match between a pair of patterns is defined here as the number of overlapping invaded pores (common to both patterns) divided by the average number of invaded pores, at breakthrough. We find an average match of 59\%, with a standard deviation of 19\%, between our simulations and experiments. 
Other metrics, including the front roughness and the main cluster saturation, are also in agreement; an exception is the Euler number---the invading phase connectivity computed as the number of clusters of air-filled pores minus the number of liquid clusters (``holes'') within them---which does not \citep{Fantinel2016}. This disparity is due to isolated liquid clusters that form and persist for longer periods in the experiments (see videos in Supporting Information). Accordingly, when considering the leading front (ignoring small isolated clusters, as often done to estimate finger width~\citep{Toussaint2012}), the pattern match improves (average of 64\% with standard deviation of 22\%).


\subsubsection{Drying Rates} 

Our simulated drying rates are generally in good agreement with experiments, except for later stages (at low saturations), where simulations exhibit fewer isolated clusters near the surface. The saturation $S$ is the ratio between the liquid volume remaining and the total pore volume. The disappearance of near-surface isolated clusters forces the drying front to recede deeper into the medium, and the simulated rate to drop (Fig.~\ref{fig:ExpSimComp}b). Here, $e$ is the evaporative flux, in terms of volume of liquid evaporated per unit time and area of open surface. 
As experimental rates are computed from the difference in the dry area between consecutive time-lapse images~\citep{Fantinel2016}, disagreement in patterns would jeopardize the match in rates. 
To further examine the impact of the drying pattern on rate, we computed the rates (using Eqs.~\ref{eq:PoreMassBal}--\ref{eq:FickPNM}) corresponding to the pore-by-pore sequence of experimental patterns. These rates match well the experimental ones, suggesting that the disparity in patterns is responsible for that in rates; it also confirms the validity of our evaporation and vapor transport calculations.

\subsection{Impact of Spatial Correlation} \label{sec:ResultsCorrImpact}
\subsubsection{Drying Patterns} \label{sec:ResultsPattern}

\begin{figure}[h!]
	\centering
	\includegraphics[width=.95\columnwidth]{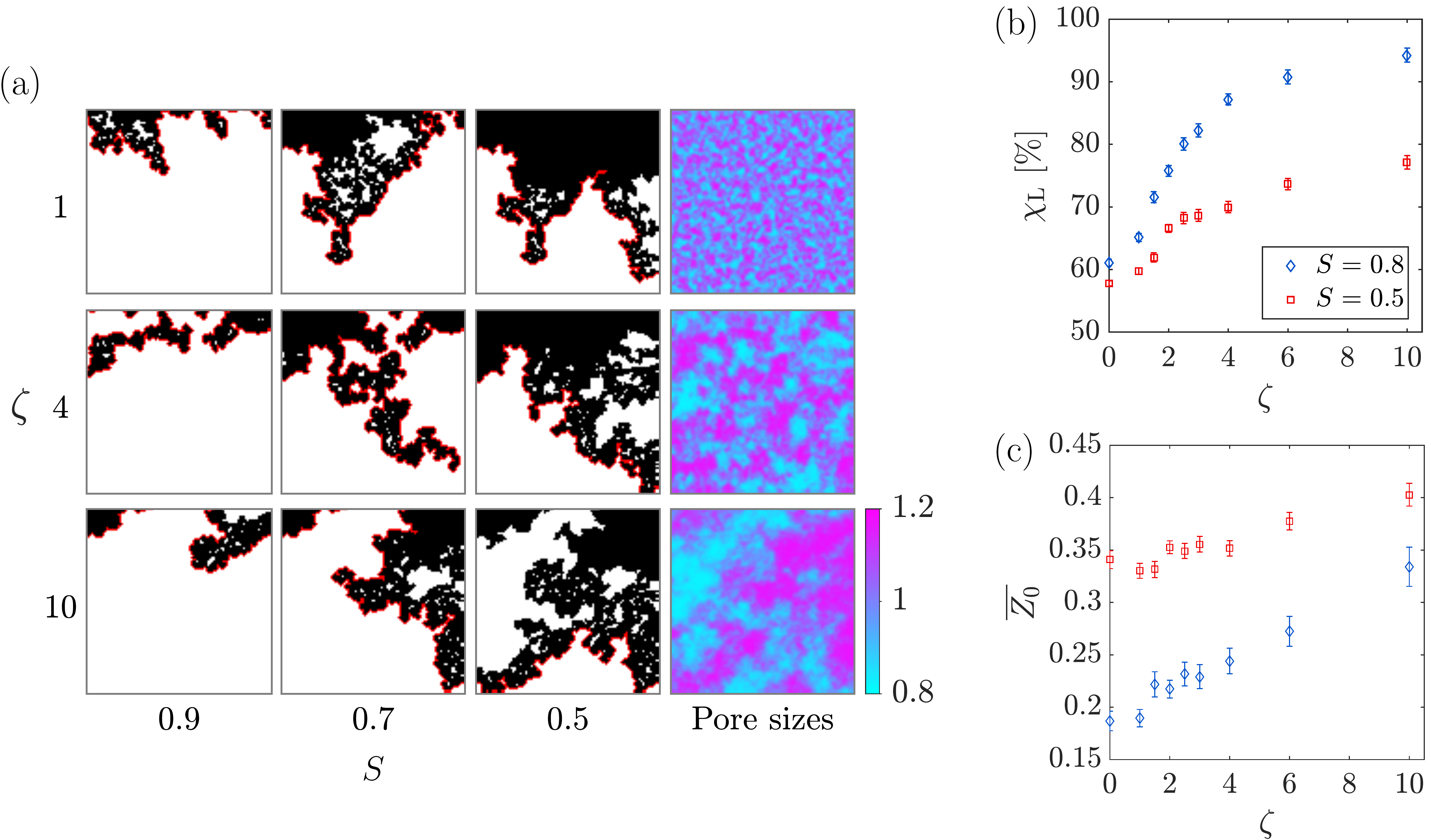}
	\caption{(a) Our simulations show that increasing correlation length $\zeta$ (in units of lattice spacing $a$), promotes preferential drying of connected large pores, such that the drying pattern follows more closely the underlying sample geometry. Patterns at different liquid saturation, $S$, are shown next to the pore size distribution (normalized by the mean).
(b) The tendency for preferential drying, quantified here through the fraction of larger-than-average invaded pores, $\chi_\mathrm{L}$ (b), and the mean invasion depth, $\overline{Z_0}$ (c), increases with $\zeta$. 
Symbols and bars represent ensemble averages and standard error from 40 realizations.}
	\label{fig:InvPattPoreSize}
\end{figure}

Our simulations demonstrate that increasing the correlation length $\zeta$ enhances  the connectivity of pores of similar size, and hence the accessibility of larger pores across the sample. This promotes preferential drying of larger pores while smaller pores remain wet and maintain liquid connectivity to the surface. 
As a result, increasing $\zeta$ forces the drying patterns to follow more closely the underlying pore geometry (Fig.~\ref{fig:InvPattPoreSize}a). Here, pore size refers to the volume between four pillars $I$, $\left( a^2 - \pi \sum_{I}R_I / 4 \right) h$.
Similarly, the tendency to invade larger pores, quantified here by the fraction of larger-than-average invaded pores (normalized by the total number of invaded pores), $\chi_\mathrm{L}$, increases with $\zeta$ (Fig.~\ref{fig:InvPattPoreSize}b). 
Another consequence of preferential drying is that it reaches deeper parts of the medium, leading to an earlier breakthrough, as demonstrated by the increase in the invasion depth $\overline{Z_0}$ (depth of the center of mass of the invaded pores) with $\zeta$ (Fig.~\ref{fig:InvPattPoreSize}c). 
%

\subsubsection{Drying Rates} \label{sec:ResInvRate}

%

\begin{figure}[h!]
	\centering
	\includegraphics[width=.7\columnwidth]{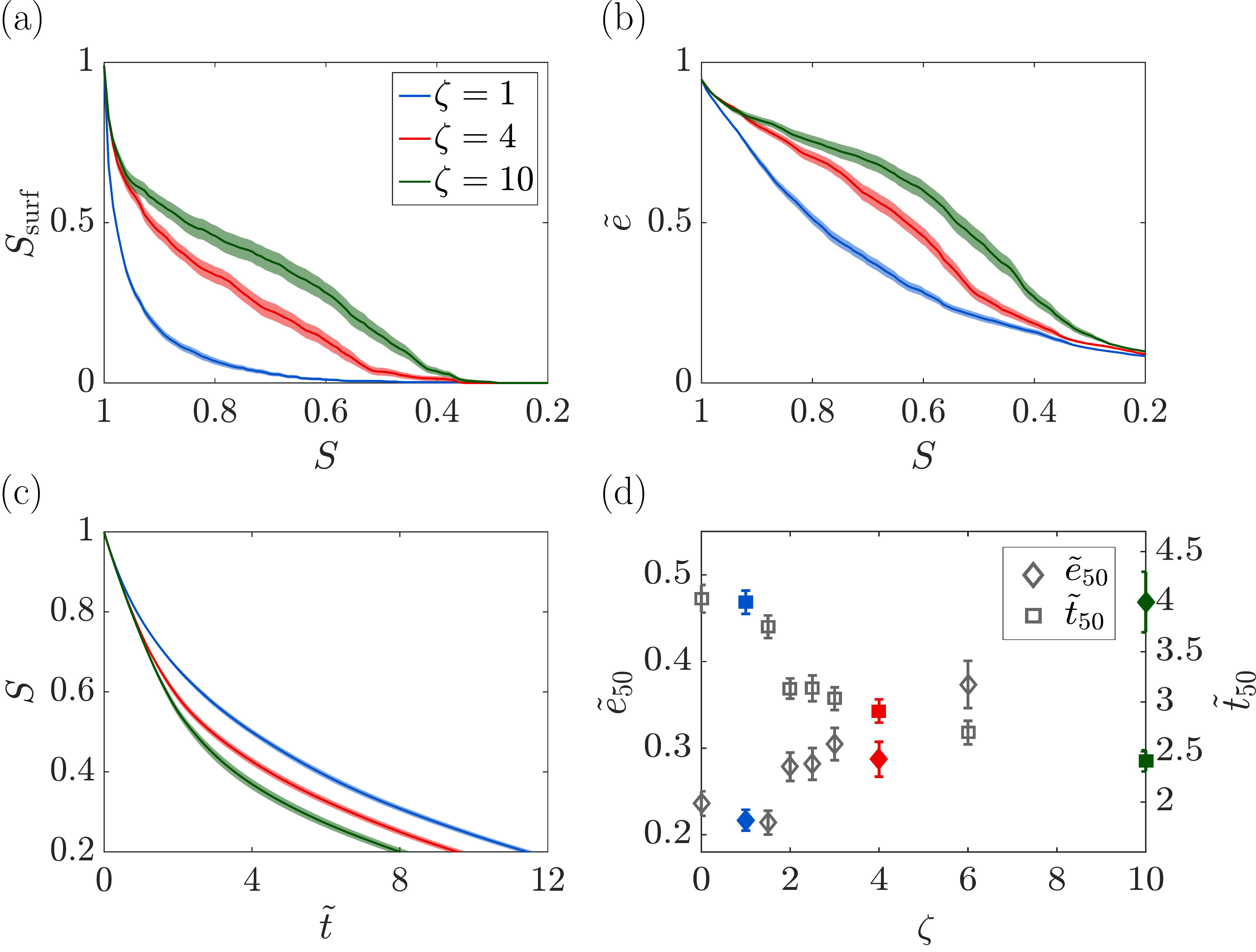}
	\caption{Increasing correlation length $\zeta$ maintains liquid connectivity to the surface, prolonging higher surface wetness, $S_\mathrm{surf}$ (a), and faster drying rates, $\tilde{e}$ (b), for longer duration, such that the saturation $S$ drops faster with time $\tilde{t}$ (c). 
	%
	Quantitatively, increasing $\zeta$ increases the rate at $S$~$=$~$50\%$, $\tilde{e}_{50}$ and decreases the time to reach $S$~$=$~$50\%$, $\tilde{t}_{50}$ (d). 
	For each $\zeta$ we plot the ensemble average (lines in a--c, symbols in d, retaining same colors) and standard error (shading in a--c, error bars in d) from 40 realizations.
	%
	}
	\label{fig:RateTimeSatMean}
\end{figure}

Maintaining liquid connectivity in more correlated samples preserves surface wetness, which, due to its strong influence on drying rates, prolongs stage 1 and delays the rate drop marking the onset of stage 2~\citep{Lehmann2008, Shokri2010a, Shahraeeni2012}. 
Our simulations capture the more gradual decrease in surface wetness, as demonstrated here via a slower drop in surface saturation $S_\mathrm{surf}$ (the liquid saturation for the row of pores closest to the boundary layer) with overall liquid saturation $S$ in samples with longer correlations $\zeta$ (Fig.~\ref{fig:RateTimeSatMean}a). This results in faster drying, as shown by: (i) maintenance of higher rates $\tilde{e}$ for longer duration (lower $S$, Fig.~\ref{fig:RateTimeSatMean}b); and (ii) faster decline of $S$ with time $\tilde{t}$ (Fig.~\ref{fig:RateTimeSatMean}c).
We use the following non-dimensional rate and time: $\tilde{e} = e/e_0$ where $e_0 = (\rho_v^\mathrm{sat} / \rho_l) D {\phi_\mathrm{sat} } / {\delta_\mathrm{BL}}$ is the potential drying rate, and $\tilde{t} = t/t_0$ where $t_0 = n L  / e_0$ is the characteristic time to evaporate liquid from a sample of depth $L$ and porosity $n$.
The enhancement of the drying rate by increasing $\zeta$ is further quantified through the larger rate at $S=50\%$, $\tilde{e}_{50}$, and shorter time to reach $S=50\%$, $\tilde{t}_{50}$ (insets of Fig.~\ref{fig:RateTimeSatMean}b--c). 
%

\section{Discussion} \label{sec:Discussion}

\subsection{Predictive Model Capabilities: Simulations vs. Experiments} \label{sec:DiscussionDiscrepancy}

Our simulated patterns are generally in good agreement with microfluidic experiments using similar pore geometry, as quantitatively shown through the agreement in pattern match, front roughness and main cluster saturation~\citep{Fantinel2016}.
The main differences between simulated and experimental patterns are the reduced formation and persistence of isolated clusters in the simulations, and the earlier rate drop compared to the experiments. 
We note that the experimental leading front (the pattern excluding isolated clusters) is well captured by our simulations. In fact, our own ongoing work shows that the leading front can be predicted by an invasion percolation (IP) model; however, we use here a more involved model since IP ignores two aspects which are crucial to the current work: (i) drying of trapped clusters and (ii) dynamics (e.g. IP, being quasi-static, ignores time which is required to determine rates). 
Potential sources for the aforementioned discrepancies include liquid films, wettability effects, and uncertainty in pore geometry; their impact is discussed below.

\subsubsection{Liquid films and wettability effects} \label{sec:Discussion_films}

Liquid films can provide a source of liquid for isolated liquid clusters close to the open surface, thus delaying invasion into these clusters and maintaining higher drying rates at lower saturations. In systems such as etched (channel) networks or granular media, such films can persist in channel corners, rough surfaces, and interstices between neighboring particles~\citep{Laurindo1998, Prat2011, Yiotis2012}.
For our pore geometry, the liquid capillary rings around solid pillars that remain after a pore is invaded are not expected to be well-connected, with little effect on liquid connectivity. This hypothesis is supported by optical microscope images from our microfluidic experiments, showing different meniscus curvatures and hence independently evolving capillary pressures in adjacent isolated liquid clusters. 

\begin{figure}
	\centering
	\includegraphics[width=.75\columnwidth]{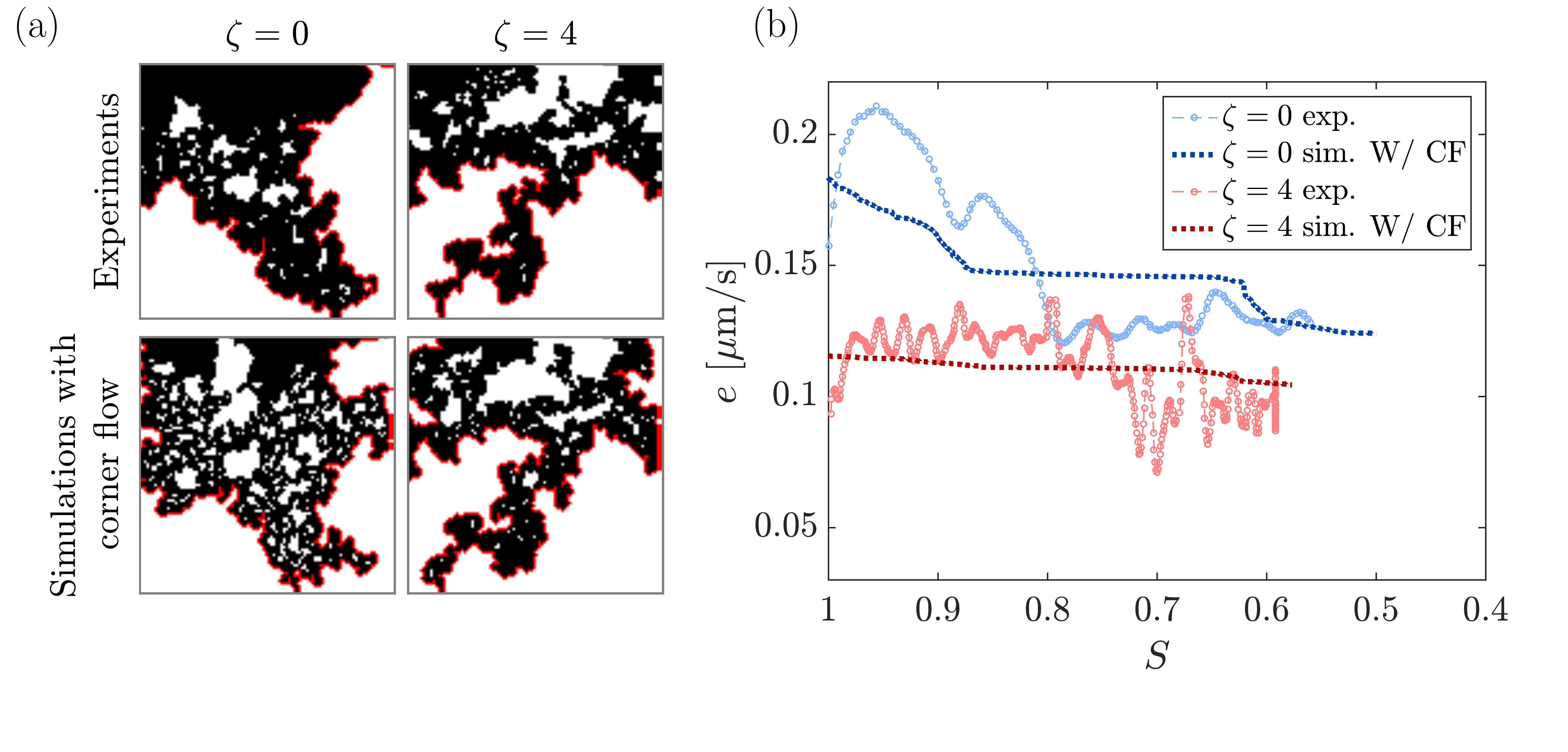}
	\caption{We evaluate the potential impact of liquid films in the corners at the cell's outer edges by simulating films which extend throughout the entire cell including the boundary layer, and during the entire experiment. Introducing these films prolongs the persistence of isolated clusters close to the open side (a) and faster drying rates (b); for comparison see $\zeta$~$=$~$0$ and $\zeta$~$=$~$4$ in Fig.~\ref{fig:ExpSimComp}. Black and white denote pores filled with air and liquid, respectively (solid not shown); the red line indicates the drying front. 
}
	\label{fig:CornFlowExpSimComp}
\end{figure}

In our experiments, however, films at the corners of the cell's edges may enhance connectivity between the main liquid reservoir and the open surface, hence promoting evaporation from depth on the expense of isolated clusters closer to the surface.  
To evaluate the potential effect of this mechanism, we simulate a scenario in which these films extend throughout the entire cell (including the part without solid pillars, i.e. the boundary layer), and during the entire experiment. 
This is crudely represented by enforcing wet pores along the cells edges (including the boundary layer cells) throughout the simulation. 
In these simulations, persistence of isolated clusters close to the open side and faster drying rates are maintained for longer periods (Fig.~\ref{fig:CornFlowExpSimComp}, compared with Fig.~\ref{fig:ExpSimComp}). 
These simulations also exhibit a more distinctive constant rate period, which could be explained by the presence of liquid films at the cell's corners. 
We stress that these simulations overestimate the effect of such films, which are expected to shrink and recede deeper into the cell during the experiment; indeed, both the isolated clusters and the high initial rates is these simulations persists longer than in the experiments (Fig.~\ref{fig:CornFlowExpSimComp}).

The reduced formation of isolated clusters could also be attributed to wettability effects. To examine the impact of wettability, we compare our experiments with highly-wetting oil (contact angle of $\sim3^{\circ}$) with one using water ($\sim70^{\circ}$). The two experiments mainly differ by the number of isolated clusters formed, suggesting that wettability effects, which are not included in our model, could decrease the agreement between our simulations and experiments~\citep{Fantinel2016}.

\subsubsection{Uncertainty in pore geometry} 

An inevitable source of disparity in patterns, and consequently in rates, is manufacturing errors (referred to here as ``noise'') introducing uncertainty in pore geometry. 
%
Our state-of-the-art manufacturing procedure provides small random errors in pillar radius, of $\sim$1.6 $\mu$m, corresponding to $\sim$3.2\% of the mean. 
The emergent patterns---in drying, and, in general, immiscible displacement---are highly sensitive to small geometrical details; that is, slight changes in pore sizes, even locally, can significantly alter the pattern~\citep{Bultreys2016}. 
An extreme example of this sensitivity is the ``binary choice'' that can occur when the invasion front reaches a bottleneck in the form of a narrow throat; if slightly altered, the invasion may proceed elsewhere, bypassing an entire region.
Such a case is presented in the Supporting Information (Fig. S3, $\lambda=0.2$, $\zeta=15$), showing distinctively different patterns in experimental samples made from the same mold (identical \textit{design}). 

\begin{figure}[h!]
	\centering
	\includegraphics[width=.5\columnwidth]{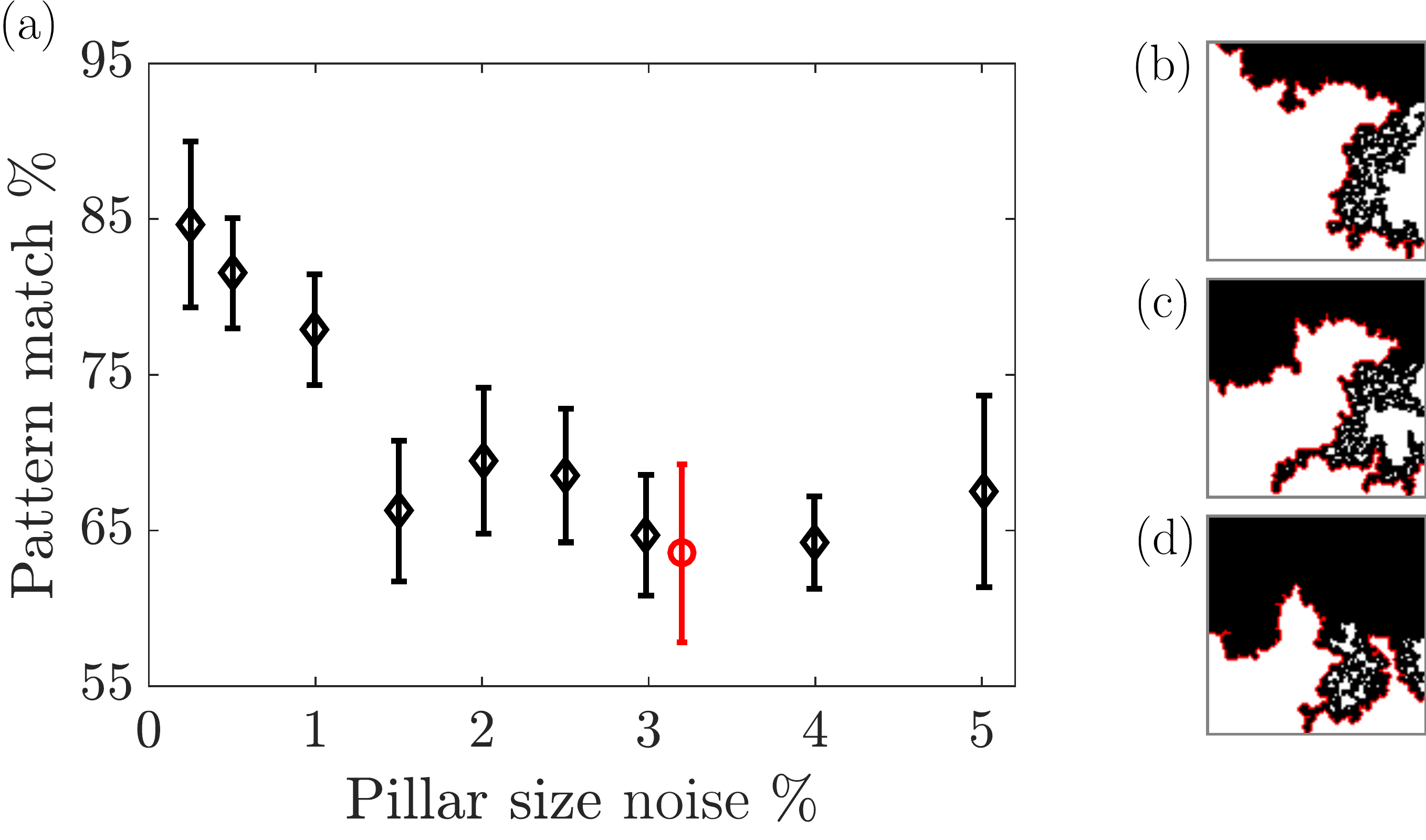}
	\caption{We evaluate the impact of geometrical uncertainty by introducing random noise in pillar sizes in the simulations, to mimic manufacturing errors. Significant changes to the emergent drying patterns, demonstrate their sensitivity to small geometrical details. (a) Our simulations show a reduction in pattern match from $\sim$85\% to $\sim$65\% (at breakthrough) as the noise increases from 0.25\% to 1.5\% (a, in black, symbols and bars are ensemble averages and standard errors from 10 samples with $\zeta$~$=$~$0$). 
	%
	Considering the observed experimental uncertainty in pillar sizes ($\sim$3.2\%), our simulated patterns agree very well with the experiments (in red, for $\zeta$~$=$~$0$). 
	The sensitivity to details is exemplified by the alteration of the drying pattern of a particular sample (b) caused by introducing an error of 1\% (c) and 3\% (d), reducing the match from 77\% to 65\%, respectively.
	%
}
	\label{fig:RndmNoiseInv}
\end{figure}

To quantify the sensitivity of the drying pattern to perturbations in pore geometry, we introduce, numerically, random noise in pillar sizes. Simulations with different noise values (applied to 10 different samples) show that the pattern match at breakthrough drops to $\sim$65\% when the error reaches 3\% (Fig.~\ref{fig:RndmNoiseInv}a). 
The fact that we obtain a comparable match between simulations and experiments indicates that our model predicts patterns very well within the experimental uncertainty in pillar sizes ($\sim$3.2\%, cf. Fig.~\ref{fig:RndmNoiseInv}a). 
To exemplify the impact of uncertainty in geometry, we show how increasing the noise introduced to a specific sample design (Fig.~\ref{fig:RndmNoiseInv}b) from 1\% to 3\% reduces the pattern match at breakthrough from 77\% to 65\% (Figs.~\ref{fig:RndmNoiseInv}c and \ref{fig:RndmNoiseInv}d, respectively).

\subsection{Impact of Correlation Length} 

\subsubsection{Pressure Evolution} %

\begin{figure}[h!]
	\centering
	\includegraphics[width=.45\columnwidth]{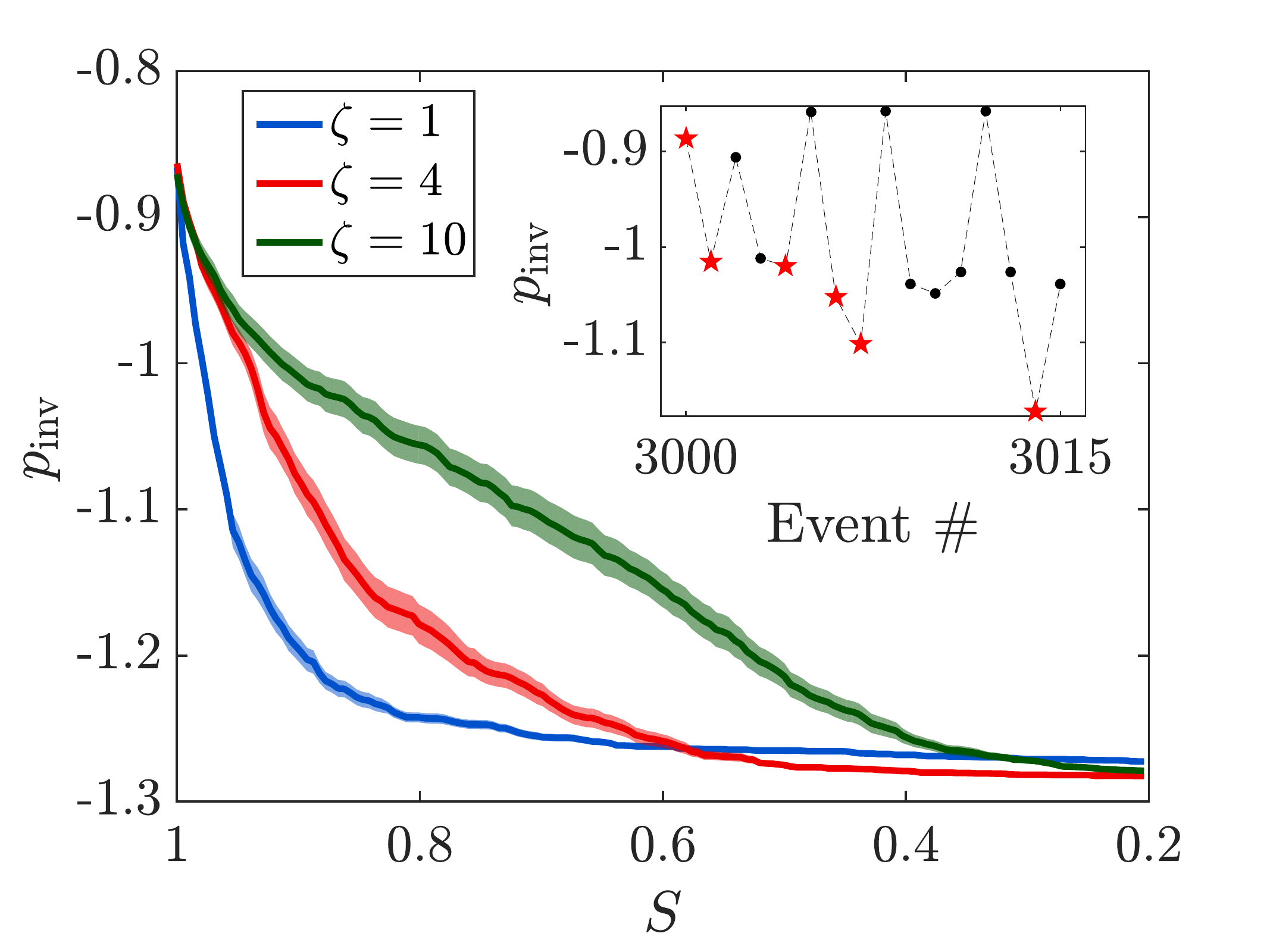}
	\caption{In more correlated media (higher $\zeta$), improved connectivity and preferential drying of large pores results in a more gradual change in the critical (minimum) invasion pressure, $p_\mathrm{min}$, as drying proceeds (with decreasing saturation, $S$). That is, a large number of pores can be invaded without a significant increase in capillary pressure (decrease in liquid pressure). 
	%
	Here $p_\mathrm{min}$ is an extreme value of the minimal pressure recorded among all liquid-filled pores during an invasion event, $p_\mathrm{inv}$, since the beginning of the simulation.
	An example of how $p_\mathrm{min}$ is computed is provided in the inset, showing the minimal liquid pressures $p_\mathrm{inv}$ for a series of 16 invasion events from a specific simulation: new (record-low) minima are marked by red stars, other events (local minima) by black dots. The main panel shows values of $p_\mathrm{min}$ vs. saturation from 40 realizations (lines and shading are ensemble average and standard error).
	 }
	\label{fig:MinInvPress}
\end{figure}

The tendency to preferentially invade large pores, which increases with correlation length, also affects the evolution of the liquid pressure. Each time the invasion front reaches a narrower throat, a further decrease in liquid pressure (by evaporation) is required to overcome the large capillary threshold. 
In a medium with non-correlated heterogeneity, the frequency of such events is high; in contrast, the interconnectivity of pores of similar sizes at high $\zeta$ reduces the frequency of such events, as exceeding a threshold enables invasion of multiple pores. 
Indeed, in samples with lower $\zeta$ we observe a sharper drop in the minimum pressure $p_\mathrm{min}$ with saturation $S$ (Fig.~\ref{fig:MinInvPress}); that is, the capillary pressure required for the front to advance becomes larger as drying proceeds and $S$ decreases, in accordance with observations from drainage simulations~\citep{Rajaram1997}. 
%
%
Here $p_\mathrm{min}$ is an extreme minimal liquid pressure recorded since the beginning of the simulations, namely a record-low value of the minimal liquid pressure during an invasion event $p_\mathrm{inv}$ (see example in inset of Fig.~\ref{fig:MinInvPress}), normalized by a characteristic invasion pressure $p^* = 2 \gamma (\overline{w}^{-1} + h^{-1})$. 

\subsubsection{Transition Between Drying Stages} 

\begin{figure}[h!]
	\centering
	\includegraphics[width=.45\columnwidth]{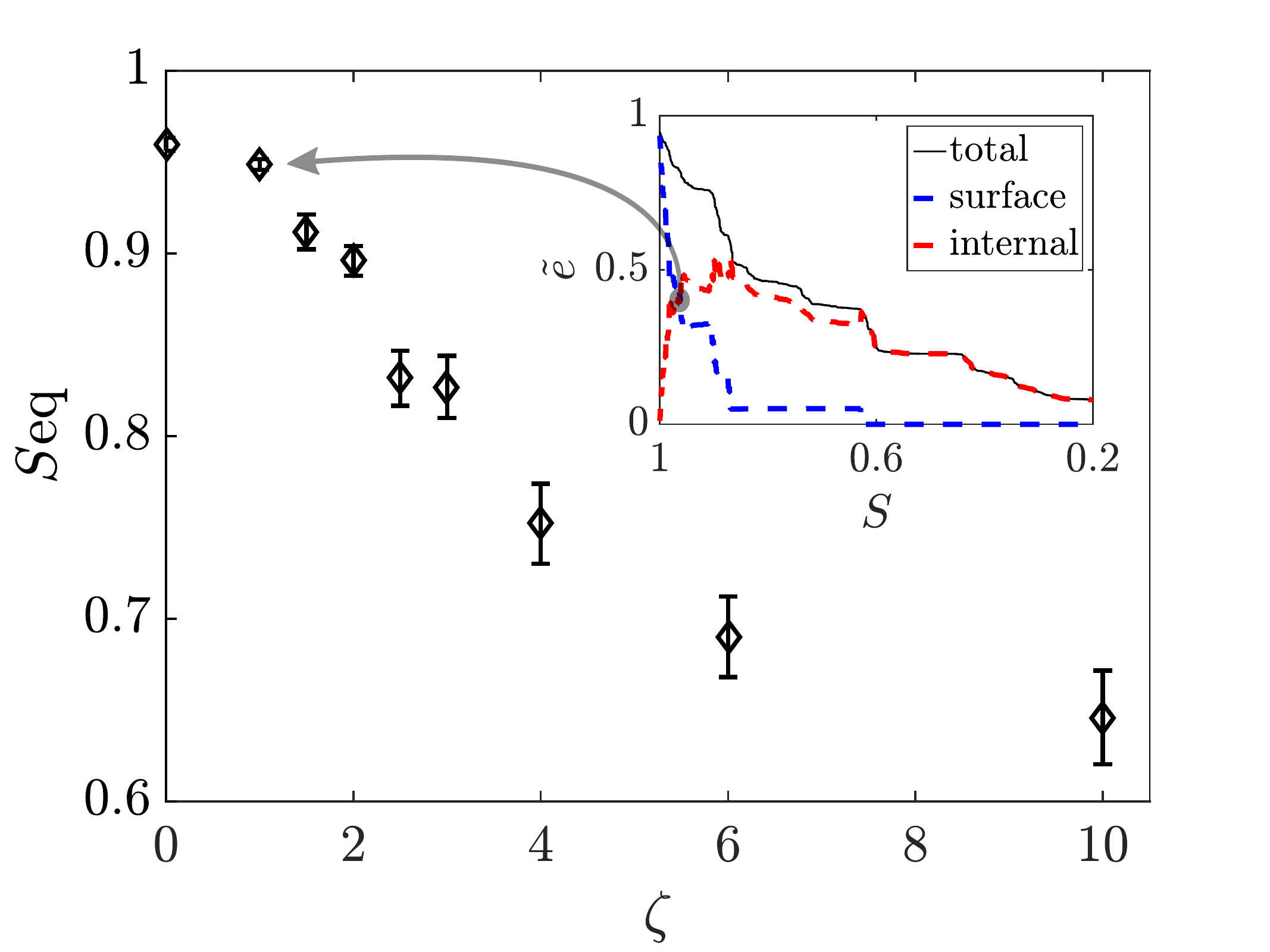}
	\caption{The transition between the drying stages is delayed as the correlation length $\zeta$ increases.
	The transition is defined here as the saturation at which the evaporation rate from surface pores equals that from internal pores, $S_\mathrm{eq}$. 
Diamonds and bars are ensemble average and standard error from 40 realizations.
The inset shows the breakup of evaporation rates for a sample with $\zeta$~$=$~1; the crossing of the red and blue lines provides $S_\mathrm{eq}$.
		 }
	\label{fig:SurfIntEvapEq}
\end{figure}

Maintaining surface wetness through liquid connectivity to the open surface controls the transition between the different drying stages, that is between dominance of evaporation from the open surface, and from deeper parts of the medium's interior~\citep{Lehmann2008}. 
Our pore-scale model allows us to quantify the relative strength of these mechanisms. We show that increasing spatial correlation delays this transition (Fig.~\ref{fig:SurfIntEvapEq}), defined here as the saturation at which the evaporation rate from the surface equals that from interior pores, $S_\mathrm{eq}$ (Fig.~\ref{fig:SurfIntEvapEq}, inset).
We note that this analysis considers connectivity of liquid-filled pores only; this transition can be further delayed by enhanced connectivity of liquid to the open surface due to film flow along channel corners~\citep{Laurindo1998,Chauvet2009} or intergranular contacts in particulate matter~\citep{Shokri2010a,Yiotis2012}.
Such a delay can be made evident in our simulations by evaluating the impact of films at the cell's outer edges
(Fig.~\ref{fig:CornFlowExpSimComp}).
The ability to predict the point of transition between drying mechanisms could be exploited in practice; for instance, one could manipulate a material's microstructure to control the duration of stage 1 drying~\citep{Assouline2014}. 


%
%

\section{Conclusions} \label{sec:Conclusions}

We study the impact of spatially-correlated pore geometry on isothermal drying of porous media. 
We present a minimal pore-scale model, describing evaporation, vapor diffusion, and capillary invasion by a set of simple rules for the interactions between pores. We compare our numerical simulations to state-of-the-art microfluidic experiments of similar pore geometry. 
Our simulated patterns compare favorably with the experiments, in light of the large sensitivity of the emergent patterns to uncertainty in pore geometry. 
We note the reduced formation and persistence of isolated clusters in simulations, leading to a faster drop in the evaporation rate than observed experimentally. Potential explanations for this discrepancy include film flow along corners at the sides of the cell, and wettability effects.

We find that increasing the correlation length promotes preferential invasion of large pores, which preserves liquid connectivity and surface wetness, maintaining higher drying rates for longer periods. 
We explain this behavior by quantifying the point of transition between dominant mechanisms (drying stages): from evaporation mostly at the surface, where rates are controlled by diffusion through the boundary layer, to evaporation from depth at a rate limited by the much slower vapor diffusion inside the porous medium.

Our approach of coupling a minimal pore-scale model with microfluidic experiments as a simple porous media analog could also be applied to study the effects of pore-scale heterogeneity in a wide range of problems including immiscible fluid-fluid displacement~\citep{Holtzman2016} and solute transport~\citep{Kang2015}. 
Specifically, our findings improve our understanding of how pore-scale heterogeneity, inevitable in most porous materials, affects their drying rate and extent. These findings bear significant implications for multiple industrial and natural processes, ranging from fuel cells~\citep{Prat2011}, cements and paints~\citep{Goehring2015} to soil-atmosphere energy and moisture exchange and soil salinization~\citep{Nachshon2011a,NorouziRad2013,Or2013a}.

\acknowledgments
Financial support by the State of Lower-Saxony, Germany (\#ZN-2823) is gratefully acknowledged.
RH also acknowledges partial support from the Israeli Science
Foundation (\#ISF-867/13) and the Israel Ministry of Agriculture
and Rural Development (\#821-0137-13).

Supporting Information includes details of the derivation of Eq. (3), drying patterns from all microfluidic experiments with corresponding simulated patterns, and videos highlighting the dynamic evolution of the drying pattern in experiments and simulations. 
The data used are available by contacting the corresponding author.


\end{document}